# Prediction of three-phase relative permeabilities of Berea sandstone using lattice Boltzmann method

Sheng Li[a], Fei Jiang[b,c,d], Bei Wei[e,f], Jian Hou[e,f], Haihu Liu[a],*

* E-mail: haihu.liu@mail.xjtu.edu.cn

[a] School of Energy and Power Engineering, Xi'an Jiaotong University,

28 West Xianning Road, Xi'an 710049, China

[b] Department of Mechanical Engineering, Yamaguchi University, Ube 7558611, Japan

[c] Blue energy center for SGE technology (BEST), Yamaguchi University, Ube 7558611,

Japan

[d] International Institute for Carbon-Neutral Energy Research (WPI-I2CNER), Kyushu

University, Fukuoka 8190395, Japan

[e] Key Laboratory of Unconventional Oil & Gas Development (China University of

Petroleum (East China)), Ministry of Education, Qingdao 266580, China.

[f] School of Petroleum Engineering, China University of Petroleum (East China),

Qingdao 266580, China

**ABSTRACT**

Three-phase flows through a pore network of Berea sandstone are studied numerically under critical interfacial tension condition. Results show that the relative permeability of each fluid increases as its own saturation increases. The specific interfacial length between wetting and non-wetting fluids monotonously decreases with increasing the saturation of intermediate-wetting fluid, while the other two specific interfacial lengths exhibit a non-monotonous variation. As the wetting (non-wetting) fluid becomes less wetting (non-wetting), the relative permeability of wetting fluid monotonously increases, while the other two relative permeabilities show a non-monotonous trend. Due to the presence of spreading layer, the specific interfacial length between wetting and non-wetting fluids always stabilizes at a low level. As the viscosity ratio of wetting (non-wetting) to intermediate-wetting fluids increases, the relative permeability of wetting (non-wetting) fluid increases. With the viscosity ratio deviating from unity, the phase interfaces become increasingly unstable, leading to an increased specific interfacial length.

## 1. Introduction

Immiscible three-phase flow through porous media has drawn much attention over the past decades, as it is commonly encountered in nature and numerous industrial processes, such as Water-Alternating-Gas (WAG) injection for enhanced oil recovery (Afzali et al., 2018; Qin et al., 2019), geological $CO_2$ sequestration (Bennion and Bachu, 2006; Jiang and Tsuji, 2016) and remediation of non-aqueous phase liquids (NAPLs) in groundwater (Javanbakht and Goual, 2016; Mccray and Falta, 1997). To predict and optimize these processes under different conditions, many relevant multiphase flow properties are required, and one of them is the relative permeability, which strongly depends on fluid properties, saturations, wettability and porous media geometry (Alizadeh and Piri, 2015) and remains poorly understood to date.

Permeability represents the capacity of a single-phase fluid to transport through the porous media, which can date back to the pioneering work of Darcy in 1856. Later, it was extended to multiphase flow by introducing the concept of relative permeability. Typically, researchers estimated three-phase relative permeabilities by extrapolating two-phase relative permeability data. On the basis of that, serval empirical correlations were proposed to predict the three-phase relative permeabilities (Baker, 1988; Blunt, 2000; Ranaee et al., 2015; Stone, 1970, 1973). However, the results obtained using these empirical correlations usually deviate significantly from the corresponding experimental data (Jiang and Tsuji, 2014; Joekar-Niasar and Hassanizadeh, 2011) due to the lack of many important mechanisms occurring in three-phase systems, such as flow coupling, double displacement, and spreading behavior of fluid (Kianinejad et al., 2016; Larsen and Skauge, 1998).

Measurement of the three-phase relative permeability can be roughly divided into two categories: the dynamic method and the static method. In the dynamic method, only one fluid is injected into the core sample to displace the in-situ fluid during the displacement process, and the fluids discharged from the sample outlet and the pressure drop across the sample are monitored at the same time. Then, the relative permeabilities can be estimated from the Johnson-Bossler-Naumann equation (Johnson et al., 1959) and its extensions (Grader and O'Meara, 1988; Virnovskii, 1984). This method is highly efficient and economical but suffers from the disadvantages that it relies on one-dimensional hypothesis and takes no account of capillary pressure, and the system may not reach the equilibrium state during the measurement. In the static method, all of the fluids remain at a fixed mass ratio within the core sample, and the relative permeabilities are calculated through multiphase Darcy's law when the system reaches the steady state. Compared to the dynamic method, the static method is time-consuming to some extent, but it is able to minimize or eliminate the non-uniform saturation distribution caused by capillary end effect so that the least controversial results can be produced. Based on these two methods, experimental studies have been conducted to measure three-phase relative permeabilities, and most of them have been summarized in the review article by Alizadeh et al. (Alizadeh and Piri, 2015). However, due to the inherent complexities of three-phase seepage and the limitations of experimental measurement, it is hard to obtain high-quality, well-characterized experimental data on three-phase relative permeabilities (Han et al., 2016). This motivates the development of high-fidelity numerical methods for the prediction of three-phase relative permeabilities, complementing the experimental studies.

Among various numerical methods, pore network (PN) models are commonly used for three-phase flow simulations in porous media (Al-Dhahli et al., 2013; Suicmez et al., 2006), which adopt the quasi-static assumption and represent the real geometry of pore space using some simplified, well-defined objects, such as spheres, tubes and so on (Blunt, 2001; Bryant and Blunt, 1992; Oren et al., 1998; Piri and Blunt, 2005). Despite their high-efficiency and popularity, the PN models still face many limitations, e.g. the effect of viscous force is usually neglected because of the quasi-static assumption (Jiang and Tsuji, 2016; Jiang et al., 2020), and the simplified geometry of pore space may neglect important pore scale phenomena (Tracy et al., 2015). Direct numerical simulations based on the realistic geometry image can be viewed as a promising option to overcome the limitations of PN models, from which one can obtain detailed information on pressure, velocity and fluid distributions. Traditionally, three-phase flows with moving contact lines (MCLs) are simulated by the volume-of-fluid method, level-set method and the phase-field method (Dong, 2017; Helland and Jettestuen, 2016; Said et al., 2014; Shi and Wang, 2014; Zhang et al., 2016). However, due to the difficulties associated to dealing with complex geometry,

MCLs and deforming interfaces with topological change, these traditional methods have not been extended to predict three-phase relative permeabilities within a realistic porous media.

The lattice Boltzmann method (LBM), as a mesoscopic method, has been developed into an alternative to traditional numerical methods for simulating complex fluid flow problems. Due to its particulate nature and local dynamics, the LBM has several advantages over traditional numerical methods, such as dealing with complex boundaries, incorporating microscopic interactions, and parallelization of the algorithm, which make it particularly suitable for the simulation of multiphase flows through complex porous media (Chen and Doolen, 1998; Xu et al., 2017). In the LBM community, several models have been proposed to simulate immiscible three-phase flows, including the color-gradient model, the pseudo-potential model, and the phase-field model. Among these models, the pseudo-potential and color-gradient models were successfully applied to simulate three-phase relative permeabilities through complex porous media (Jiang and Tsuji, 2016; Zhang et al., 2019). However, due to the limitation of the models, these works are limited to either the case of totally wetting condition in which one fluid is totally wetting to the rock surface, or the case in which three interfacial tensions satisfy a Neumann's triangle. This is evidently different from the situation in a realistic ternary fluid system, e.g. the WAG injection process, where three interfacial tensions often do not satisfy the Neumann's triangle (Egermann et al., 2000; Element et al., 2003; Oak, 1990), and the rock may be oil-wet, water-wet or mixed wet, with the contact angles varying over a wide range (Anderson, 1986; Xu et al., 2020). In addition, the influence of many factors, including the fluid saturations, surface wettability and viscosity ratios, under realistic conditions, and the corresponding pore-scale mechanisms remain unclear so far (Jiang and Tsuji, 2016;

Zhang et al., 2019).

To fill the gap mentioned above, we will use the lattice Boltzmann color-gradient model recently improved by our group (Li et al., 2021; Yu et al., 2019b) to simulate the immiscible three-phase flows in a two-dimensional (2D) pore network of realistic Berea sandstone. In particular, we will focus on a ternary fluid system where the largest interfacial tension equals the sum of the other two. By applying the static method in which the fluid movement is driven by a constant body force, the effect of fluid saturation, surface wettability and viscosity ratio on the steady-state fluid distribution, relative permeability, and the specific interfacial length are systematically investigated and analyzed. The rest of the paper is organized as follows. The numerical method and validation are given in the next section. Then, the problem description and the influence of various parameters are presented and discussed in Section 3. Finally, the paper closes with a summary of the main results in Section 4.

## 2. Numerical method and validation

*2.1. Lattice Boltzmann color-gradient model for immiscible three-phase flows*

A LB color-gradient model recently developed by Yu et al. (Yu et al., 2019b) is used for the pore-scale simulation of immiscible three-phase flows within porous media. In this model, three different fluids, namely blue, green and red fluids, are considered, and their distribution functions are denoted as $f_{\alpha,b}$, $f_{\alpha,g}$ and $f_{\alpha,r}$, respectively. The implementation of this model consists of three steps, i.e. collision, recoloring and streaming, in each time step. First, the collision step is implemented for the total distribution function, which reads as

$$f_\alpha^\dagger(\mathbf{x},t) = f_\alpha(\mathbf{x},t) + \Omega_\alpha(\mathbf{x},t) + \overline{F}_\alpha(\mathbf{x},t), \tag{1}$$

where $f_\alpha(\mathbf{x},t)$ is the total distribution function in the $\alpha$ th velocity direction at the position $\mathbf{x}$ and time $t$ and it is defined as $f_\alpha = \sum_k f_{\alpha,k}$ ($k = b$, $g$ or $r$); $f_\alpha^\dagger$ is the post-collision distribution function; $\Omega_\alpha$ is the collision operator and $\bar{F}_\alpha$ is the forcing term.

To improve the model stability for high viscosity ratio problems and produce viscosity-independent permeability in porous media, the multiple-relaxation-time (MRT) collision operator is adopted, which is given by (Pan et al., 2004; Pan et al., 2006)

$$\Omega_\alpha(\mathbf{x},t) = -\sum_\beta \left(\mathbf{M}^{-1}\mathbf{S}\mathbf{M}\right)_{\alpha\beta} \left[f_\beta(\mathbf{x},t) - f_\beta^{eq}(\mathbf{x},t)\right], \qquad (2)$$

where $\mathbf{M}$ is the transformation matrix and $\mathbf{S}$ is a diagonal relaxation matrix. $f_\alpha^{eq}$ is the equilibrium distribution function of $f_\alpha$, and is defined by the local fluid velocity $\mathbf{u}$ as

$$f_\alpha^{eq}(\mathbf{x},t) = \rho w_\alpha \left[1 + \frac{\mathbf{e}_\alpha \cdot \mathbf{u}}{c_s^2} + \frac{(\mathbf{e}_\alpha \cdot \mathbf{u})^2}{2c_s^4} - \frac{\mathbf{u} \cdot \mathbf{u}}{2c_s^2}\right], \qquad (3)$$

where the total density $\rho$ is calculated by $\rho = \sum_k \rho_k$ with $\rho_k$ referring to the density of the fluid $k$; $c_s = \delta_x/(\sqrt{3}\delta_t) = 1/\sqrt{3}$ is the speed of sound with the lattice spacing $\delta_x$ and the time step $\delta_t$ both taken as 1; $w_\alpha$ is the weight coefficient, and $\mathbf{e}_\alpha$ is the lattice velocity in the $\alpha$ th direction. For the two-dimensional nine-velocity (D2Q9) lattice model used in the present study, the lattice velocity $\mathbf{e}_\alpha$ is defined as $\mathbf{e}_0 = (0,0)$, $\mathbf{e}_{1,3} = (\pm 1, 0)$, $\mathbf{e}_{2,4} = (0, \pm 1)$, $\mathbf{e}_{5,6} = (\pm 1, 1)$, $\mathbf{e}_{7,8} = (\mp 1, -1)$, and the weight coefficient $w_\alpha$ is given by $w_0 = 4/9$, $w_{1-4} = 1/9$, $w_{5-6} = 1/36$.

The transformation matrix $\mathbf{M}$ and the diagonal relaxation matrix $\mathbf{S}$ are given by (Lallemand and Luo, 2000; Luo et al., 2011)

$$\mathbf{M} = \begin{pmatrix} 1 & 1 & 1 & 1 & 1 & 1 & 1 & 1 & 1 \\ -4 & -1 & -1 & -1 & -1 & 2 & 2 & 2 & 2 \\ 4 & -2 & -2 & -2 & -2 & 1 & 1 & 1 & 1 \\ 0 & 1 & 0 & -1 & 0 & 1 & -1 & -1 & 1 \\ 0 & -2 & 0 & 2 & 0 & 1 & -1 & -1 & 1 \\ 0 & 0 & 1 & 0 & -1 & 1 & 1 & -1 & -1 \\ 0 & 0 & -2 & 0 & 2 & 1 & 1 & -1 & -1 \\ 0 & 1 & -1 & 1 & -1 & 0 & 0 & 0 & 0 \\ 0 & 0 & 0 & 0 & 0 & 1 & -1 & 1 & -1 \end{pmatrix} \tag{4}$$

and

$$\mathbf{S} = \mathrm{diag}\left[0,\ \omega,\ \omega,\ 0,\ \frac{8(2-\omega)}{8-\omega},\ 0,\ \frac{8(2-\omega)}{8-\omega},\ \omega,\ \omega\right], \tag{5}$$

where $\omega$ is the dimensionless relaxation parameter related to the dynamic viscosity $\mu$ of the fluid mixture by $\mu = (1/\omega - 0.5)\rho c_s^2 \delta_t$. For the sake of simplicity, three immiscible fluids are assumed to have equal densities. To allow for unequal viscosities for these fluids, a harmonic mean is adopted to determine the viscosity of the fluid mixture, i.e.,

$$\frac{\rho}{\mu} = \sum_k \frac{\rho_k}{\mu_k}, \tag{6}$$

where $\mu_k$ is the dynamic viscosity of the pure fluid $k$.

The forcing term $\bar{F}_\alpha$ is included to realize the effects of interfacial tension and body force, which reads as (Yu and Fan, 2010)

$$\bar{\mathbf{F}} = \mathbf{M}^{-1}\left(\mathbf{I} - \frac{1}{2}\mathbf{S}\right)\mathbf{M}\tilde{\mathbf{F}}, \tag{7}$$

where $\mathbf{I}$ is a $9\times 9$ unit matrix, $\bar{\mathbf{F}} = [\bar{F}_0,\ \bar{F}_1,\ \bar{F}_2,\ \cdots,\ \bar{F}_8]^T$, and $\tilde{\boldsymbol{F}} = [\tilde{F}_0,\ \tilde{F}_1,\ \tilde{F}_2,\ \cdots,\ \tilde{F}_8]^T$ with the element $\tilde{F}_\alpha$ given by (Guo et al., 2002)

$$\tilde{F}_\alpha = w_\alpha \left[ \frac{\mathbf{e}_\alpha - \mathbf{u}}{c_s^2} + \frac{(\mathbf{e}_\alpha \cdot \mathbf{u})\mathbf{e}_\alpha}{c_s^4} \right] (\boldsymbol{F}_s + \boldsymbol{F}_b)\delta_t. \tag{8}$$

Herein, $\boldsymbol{F}_b$ is a body force, and $\boldsymbol{F}_s$ is the interfacial tension force expressed as (Yu et al., 2019b)

$$\boldsymbol{F}_s = \sum_k \sum_{l,l \neq k} \nabla \cdot \left[ \frac{\sigma_{kl} A_{kl}}{2} |\mathbf{G}_{kl}| (\mathbf{I} - \mathbf{n}_{kl}\mathbf{n}_{kl}) \right], \tag{9}$$

where $\mathbf{G}_{kl} = C_l \nabla C_k - C_k \nabla C_l$ is the color gradient parameter and $C_k = \rho_k / \rho$ is the local mass fraction of the fluid $k$. $\sigma_{kl}$ is the interfacial tension between the fluid $k$ and the fluid $l$, and $\mathbf{n}_{kl} = \mathbf{G}_{kl} / |\mathbf{G}_{kl}|$ is the unit normal vector of the $k$-$l$ interface. $A_{kl}$ is a concentration factor given by $A_{kl} = \min\left(10^6 \rho_k \rho_l / (\rho_k^0 \rho_l^0), 1\right)$, where $\rho_k^0$ is the density of the pure fluid $k$ (Leclaire et al., 2013). The fluid velocity $\mathbf{u}$ is defined as

$$\rho \mathbf{u}(\mathbf{x},t) = \sum_\alpha f_\alpha(\mathbf{x},t)\mathbf{e}_\alpha + \frac{1}{2}[\boldsymbol{F}_s(\mathbf{x},t) + \boldsymbol{F}_b(\mathbf{x},t)]\delta_t \tag{10}$$

to exactly recover the Navier-Stokes equations in the low frequency, long wavelength limit.

To produce the phase segregation and maintain a reasonable interface, a recoloring step is then applied. Following the recoloring algorithm of Spencer et al. (Spencer et al., 2010), the recolored distribution function of the fluid $k$ is given by

$$f_{\alpha,k}^{\dagger\dagger}(\mathbf{x},t) = \frac{\rho_k}{\rho} f_\alpha^\dagger(\mathbf{x},t) + \sum_{l,l \neq k} \beta_{kl} \omega_\alpha \frac{\rho_k \rho_l}{\rho} \mathbf{n}_{kl} \cdot \mathbf{e}_\alpha, \tag{11}$$

where $f_{\alpha,k}^{\dagger\dagger}$ is the recolored distribution function of the fluid $k$. $\beta_{kl}$ is the segregation parameter defined as (Yu et al., 2019a)

$$\beta_{kl} = \beta^0 + \beta^0 \min\left(\frac{35\rho_b\rho_g\rho_r}{\rho^3}, 1\right) g(X_{kl}), \tag{12}$$

with

$$g(X_{kl}) = \begin{cases} 1, & X_{kl} < -1 \\ 1 - \sin(\arccos(X_{kl})), & -1 \leq X_{kl} < 0 \\ \sin(\arccos(X_{kl})) - 1, & 0 \leq X_{kl} \leq 1 \\ -1, & 1 < X_{kl} \end{cases}. \tag{13}$$

Where the reference segregation parameter $\beta^0$ is fixed at 0.7 (Halliday et al., 2007; Liu et al., 2012), and $X_{kl} = (\sigma_{mk}^2 + \sigma_{ml}^2 - \sigma_{kl}^2)/(2\sigma_{mk}\sigma_{ml})$. It has been demonstrated that the recoloring algorithm given by Eqs.(11)-(13) is able to accurately simulate three-phase flows with a full range of interfacial tensions especially at the critical and super-critical states, where the largest interfacial tension equals or exceeds the sum of the other two (Yu et al., 2019b).

After the recoloring step, the distribution function of the fluid $k$ propagates to the neighboring lattice nodes, known as the streaming step, which reads as

$$f_{\alpha,k}(\mathbf{x} + \mathbf{e}_\alpha \delta_t, t + \delta_t) = f_{\alpha,k}^{\dagger\dagger}(\mathbf{x}, t), \tag{14}$$

and the resulting distribution functions are used to compute the densities of three immiscible fluids, i.e., $\rho_k = \sum_\alpha f_{\alpha,k}$.

*2.2. Wetting boundary condition*

The wettability of solid surfaces can be described by three equilibrium contact angles, i.e., $\theta_{rg}$, $\theta_{br}$ and $\theta_{gb}$. According to the Young's equation, when three interfacial tensions are specified, these three equilibrium contact angles cannot be arbitrarily chosen, and they must satisfy the following relationship (Yu et al., 2019a;

Zhang et al., 2016):

$$\sigma_{rg}\cos\theta_{rg}+\sigma_{br}\cos\theta_{br}+\sigma_{gb}\cos\theta_{gb}=0, \qquad (15)$$

where $\theta_{kl}$ is the contact angle made by the $k$-$l$ interface with the solid surface, and is measured from the side of the fluid $k$. It is noted that the contact angles $\theta_{kl}$ and $\theta_{lk}$ are a pair of complementary angles, i.e., $\theta_{lk}=\pi-\theta_{kl}$.

The wetting boundary condition is implemented using the characteristic line method recently proposed by Li et al. (Li et al., 2021), in which the desired contact angles are realized by specifying the values of mass fraction at the solid nodes adjacent to the boundary (a list of these nodes is denoted as $\Omega_{S_B}$) using two possible characteristic lines. The implementation of the boundary condition will be briefly described below, and one can refer to Li et al. (Li et al., 2021) for more details.

Firstly, we evaluate the value of $C_k$ in $\Omega_{S_B}$ through a weighted average of its nearest $C_k$ in the fluid nodes adjacent to the boundary (similarly, a list of these nodes is denoted as $\Omega_{F_B}$), which is given by (Xu et al., 2017)

$$C_k(\mathbf{x})=\frac{\sum_{(i:\mathbf{x}+\mathbf{e}_i\delta_t\in\Omega_{F_B})}w_i C_k(\mathbf{x}+\mathbf{e}_i\delta_t)}{\sum_{(i:\mathbf{x}+\mathbf{e}_i\delta_t\in\Omega_{F_B})}w_i}, \quad \mathbf{x}\in\Omega_{S_B}. \qquad (16)$$

Secondly, we use the weighted contact angle method proposed by Zhang et al. (Zhang et al., 2016) to determine the local contact angles $\theta_r$ and $\theta_g$ in $\Omega_{S_B}$ from the evaluated $C_k$, which can be expressed as

$$\begin{cases}\theta_r=\dfrac{C_b}{1-C_r}\theta_{rb}+\dfrac{C_g}{1-C_r}\theta_{rg}\\[6pt]\theta_g=\dfrac{C_b}{1-C_g}\theta_{gb}+\dfrac{C_r}{1-C_g}\theta_{gr}\end{cases}. \qquad (17)$$

Thirdly, for each node in $\Omega_{S_B}$, we find the characteristic lines originating from it according to the obtained contact angles $\theta_r$ and $\theta_g$. Note that two characteristic lines i.e., $l_1$ and $l_2$, can be obtained for a specific $\theta_k$ ($k=r$ or $g$), which make an included angle of $\left|\frac{\pi}{2}-\theta_k\right|$ with the unit normal vector of the solid surface $\mathbf{n}_s$ and are symmetric with respect to $\mathbf{n}_s$.

Fourthly, we extend the characteristic lines $l_1$ and $l_2$ until they intersect the grid lines at $D_1$ and $D_2$, respectively. Then, we can compute the values of $C_k$ at the points $D_1$ and $D_2$, i.e. $C_{D_1}$ and $C_{D_2}$, through a linear interpolation of $C_k$ at their surrounding grid nodes. After obtaining $C_{D_1}$ and $C_{D_2}$ we choose and assign one of them to the corresponding solid node using the following discriminant function (Liu and Ding, 2015)

$$C_P = \begin{cases} \max\{C_{D1}, C_{D2}\} & \theta_k \leq 90°; \\ \min\{C_{D1}, C_{D2}\} & \theta_k > 90°. \end{cases} \qquad (18)$$

Once the above treatment is applied to all the solid nodes in $\Omega_{S_B}$, the desired contact angles are implicitly imposed by the gradient of $C_k$ in $\Omega_{F_B}$.

*2.3. Validation*

In our previous work (Li et al., 2021), the present model has been validated by a series of static and dynamic tests, and it was demonstrated to be capable of accurately predicting three-phase flows with moving contact lines on arbitrarily complex solid surfaces. Here, we use this model to simulate the layered three-phase flow in a channel to assess whether the body force $\mathbf{F}_b$ is properly incorporated.

The geometry setup of the problem is shown in Figure 1. Three fluids are distributed symmetrically with respect to the center line in the channel. Specifically, the blue fluid is located in the central region of $|y|<a$, green fluid in the region of $a<|y|<b$, and the red fluid is distributed in the region of $b<|y|<H$. Periodic boundary conditions are used at the left and right boundaries, while the halfway bounce-back scheme is applied at the bottom and top walls. A constant body force $\boldsymbol{F}_b$ along the x-direction is applied to drive the fluid flow in the channel. Assuming a Poiseuille-type flow in the channel, the analytical solution for the velocity profile $u(y)$ can be obtained (Jiang et al., 2020)

$$u(y) = \begin{cases} A_1 y^2 + C_1 & 0 \leq y < a \\ A_2 y^2 + B_2 y + C_2 & a \leq y < b \\ A_3 y^2 + B_3 y + C_3 & b \leq y \leq H \end{cases}, \tag{19}$$

where

$$A_1 = 0.5 F_b / \mu_b, \quad A_2 = 0.5 F_b / \mu_g, \quad A_3 = 0.5 F_b / \mu_r, \tag{20}$$

$$B_2 = -2A_2 a + 2A_1 a \mu_b / \mu_g, \tag{21}$$

$$B_3 = -2A_3 b + (2A_2 b + B_2) \mu_g / \mu_r, \tag{22}$$

$$C_3 = -A_3 H^2 - B_3 H, \tag{23}$$

$$C_2 = (A_3 - A_2) b^2 + (B_3 - B_2) b + C_3, \tag{24}$$

$$C_1 = (A_2 - A_1) a^2 + B_2 b + C_2. \tag{25}$$

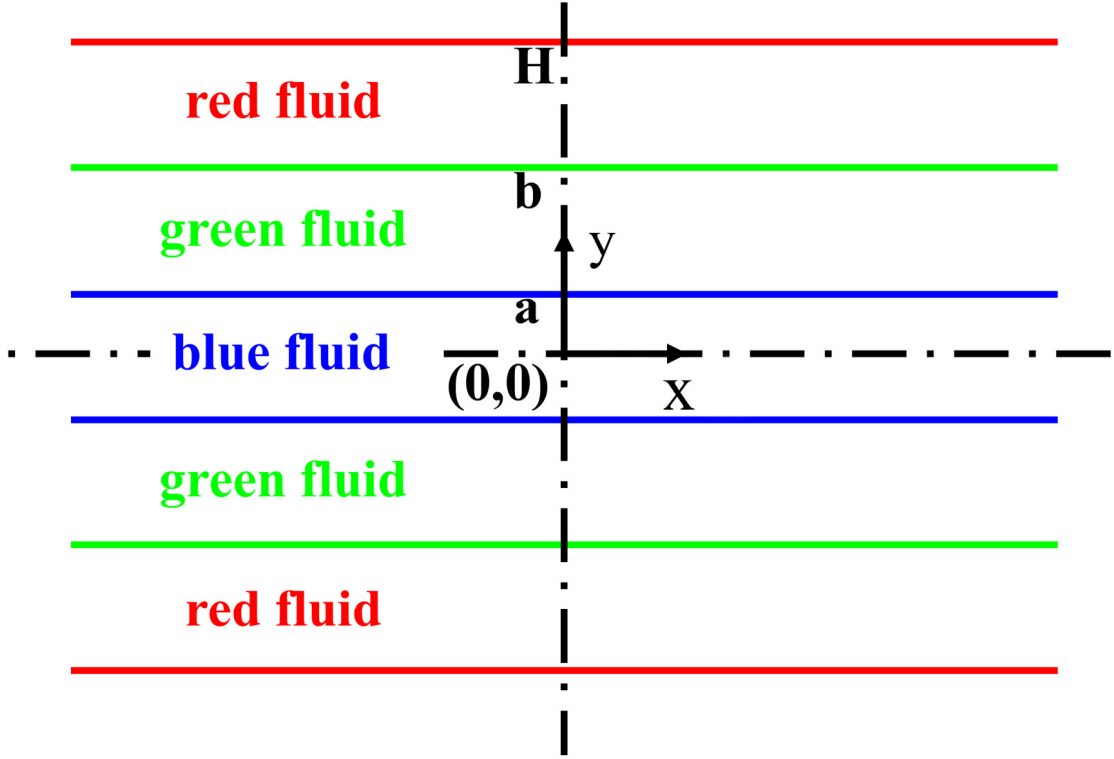

**FIG. 1.** Schematic of the layered three-phase flow in a 2D channel. The blue fluid flows in the central region $|y|<a$, the green fluid in the region $a<|y|<b$, and the red fluid flows in the region $b<|y|<H$.

In the simulations, the size of the computational domain is set as $15\times 250$ lu$^2$ (lu represents the lattice length unit), and five layers of fluid have the same height of 50 lu. The magnitude of the body force $F_b$ is $5\times 10^{-8}$, and the maximum viscosity among three fluids is fixed at 0.25. Two group of viscosity ratios are considered: (a) $\mu_b:\mu_g:\mu_r = 25:5:1$ and (b) $\mu_b:\mu_g:\mu_r = 1:5:25$. Figure 2 shows the comparison between simulated velocity profiles and the analytical solutions for the two cases, and the perfect agreement is obtained, indicating that the body force term has been incorporated into the LBM model correctly. By contrast, the numerical results obtained from the three-component pseudo-potential LBM noticeably deviate from the analytical solutions especially in the vicinity of interfaces (Wei et al., 2018).

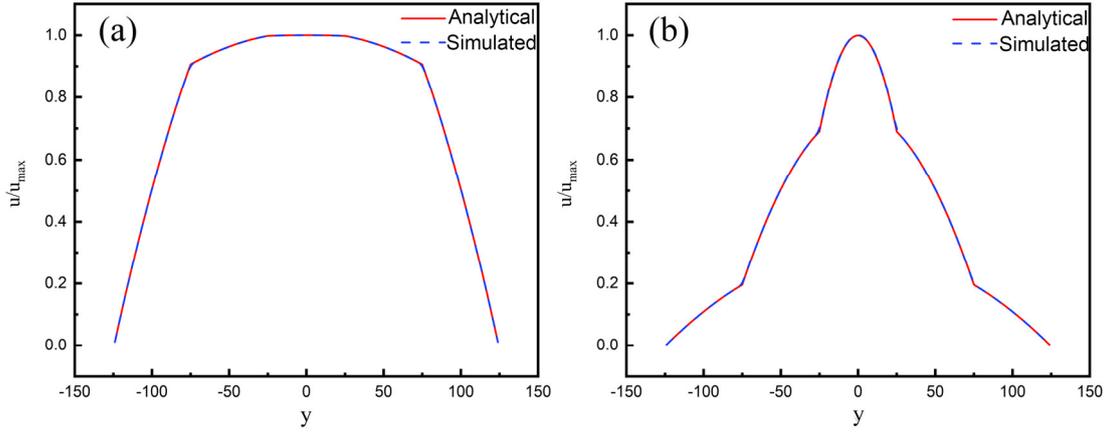

**FIG. 2.** Comparison of the velocity profiles obtained by the present model and the analytical solutions at the viscosity ratios of (a) $\mu_b : \mu_g : \mu_r = 25:5:1$ and (b) $\mu_b : \mu_g : \mu_r = 1:5:25$. Note that the fluid velocity $u$ is normalized by its maximal value $u_{max}$. The analytical solutions and the simulated results are represented by the red solid lines and the blue dashed lines, respectively.

## 3. Results and discussion

In this section, the immiscible three-phase flows driven by a constant body force within a 2D pore network are simulated. At first, we give a brief description about this problem. Then, the effect of fluid saturations, surface wettability and viscosity ratios is systematically studied and discussed.

### 3.1. Description of the problem

As shown in Figure 3, a 2D pore network extracted from a slice of realistic Berea sandstone is adopted for the simulations. The pore network has a porosity of 30% and its intrinsic permeability normalized by the area of pore network is 3.261. In the simulations, it is discretized into a domain of $L_x \times L_y = 1500 \times 1300$ lu$^2$. The halfway bounce-back scheme and the wetting boundary condition are imposed at the rock surfaces to achieve the desired contact angles. Periodic boundary conditions are used at all sides of the domain, which allows a constant saturation for each fluid during the simulation. It is also noted that five layers of ghost nodes are added to each side of the domain to ensure the connectivity when periodic boundary conditions are applied, so the size of the actual computational domain is $1510 \times 1310$ lu$^2$. However, the fluids in the ghost layers are excluded when calculating the statistical quantities like the saturations and fluxes.

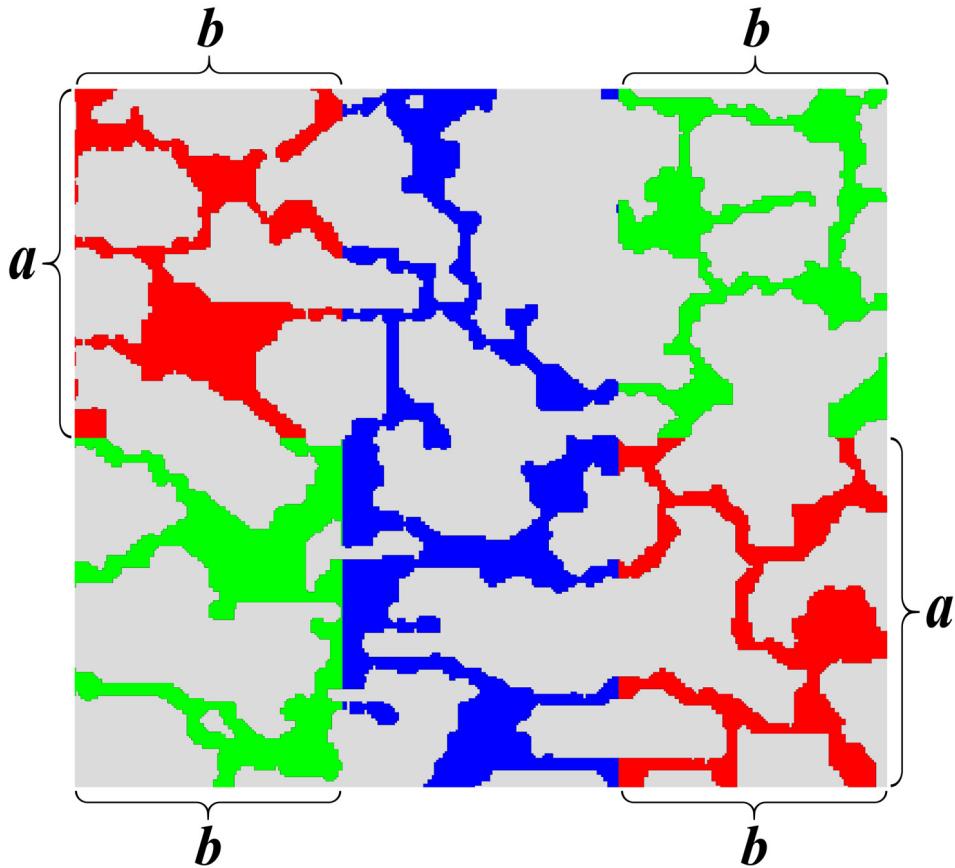

**FIG. 3.** The pore network generated from an image of realistic Berea sandstone (Boek and Venturoli, 2010), and the initial fluid distributions within it. Note that the solid grains are represented in gray, while the void space occupied by three fluids is indicated in blue, green or red.

Figure 3 shows how the fluid saturations $S_b$, $S_g$ and $S_r$ are initialized in the simulation. Initially, the blue fluid is distributed in the middle part of the computational domain with the size of $(L_x - 2b) \times L_y$; the other two fluids occupy four corners with the red fluid in the left-top and right-bottom corners and the green fluid in the left-bottom and right-top corners. Different values of $S_b$, $S_g$ and $S_r$ can be achieved by adjusting the parameter $a$ or $b$. A constant body force $\boldsymbol{F}_b = (F_b, 0)$ along the positive *x*-direction is used to drive the fluid movement, which leads to a capillary number of (Huang and Lu, 2009; Yiotis et al., 2007)

$$Ca = \frac{F_b}{\sigma_{bg}}. \tag{26}$$

Unless otherwise stated, the capillary number of the system is fixed at $Ca = 5.0 \times 10^{-4}$, and the interfacial tension ratio is set to $\sigma_{gr} : \sigma_{bg} : \sigma_{br} = 2:1:1$ as the critical and supercritical states are often encountered in realistic applications, e.g. Water-Alternating-Gas injection (Egermann et al., 2000; Element et al., 2003; Oak, 1990). When the simulation reaches the steady state, the Darcy velocities and the relative permeabilities of three fluids are calculated by

$$u_i = \frac{\sum_{\rho_i(\mathbf{x})>0} u_x(\mathbf{x})}{L_x \times L_y}, \quad i = b,\ g,\ r \tag{27}$$

and

$$K_i = \frac{u_i \mu_i / F_b}{k_\infty}, \quad i = b,\ g,\ r, \tag{28}$$

where $u_x$ is the x-component of the velocity $\boldsymbol{u}$, and $k_\infty$ is the intrinsic permeability of the pore network. Another important parameter in three-phase flow through porous media is the specific interfacial length, which is closely related to the mass and energy transfer between fluids. In this study, the specific interfacial length $L_{kl}$ is defined as the ratio of the interfacial length between fluids $k$ and $l$ to the pore area.

*3.2. Effect of fluid saturations*

We first investigate the effect of the saturations on the relative permeabilities for three fluids with equal viscosities. The contact angles are set to $\theta_{bg} = 45°$, $\theta_{br} = 135°$ and $\theta_{gr} = 135°$, meaning that the rock surface is wetting to the red fluid, intermediate wetting to the blue fluid, and non-wetting to the green fluid. Seven cases with different groups of fluid saturations are simulated. Note that the saturations of red and green fluids are kept roughly equal when changing the saturations, which is realized by adjusting the parameter $b$ whilst keeping the parameter $a$ as a constant in the initial state. Figure 4 plots the relative permeability $K_i$ as a function of saturation $S_i$ for three different fluids. It can be seen that for each fluid, the relative permeability increases with its saturation. This can be explained by the fluid distributions shown in Figure 5, where three cases are selected as representatives, i.e., (a) $S_b = 73.22\%$, $S_g = 11.51\%$, $S_r = 15.27\%$, (b) $S_b = 46.09\%$, $S_g = 25.79\%$, $S_r = 28.12\%$, and (c) $S_b = 19.63\%$, $S_g = 41.74\%$, $S_r = 38.63\%$. It is clear that the fluid with lower saturation always shows a scattered and disconnected distribution in the void space, resulting in a poor ability to flow through the porous media, especially for the wetting fluid which tends to adhere to the rock surface or be trapped in the small pores. Thus, it is expected that the relative permeability is small when the saturation is low. On the contrary, when the saturation is high, the fluid is distributed continuously in the void space and forms active flow paths along the horizontal direction, leading to a large relative permeability. Such a trend of relative permeability $K_i$ versus saturation $S_i$ was also reported in previous theoretical and numerical studies (Jiang and Tsuji, 2016; Stone, 1970). Quantitatively, the relationship between saturation and relative permeability for the wetting phase (red fluid) exhibits a nonlinear behavior, which is in agreement with the experimental data (Qin et al., 2018).

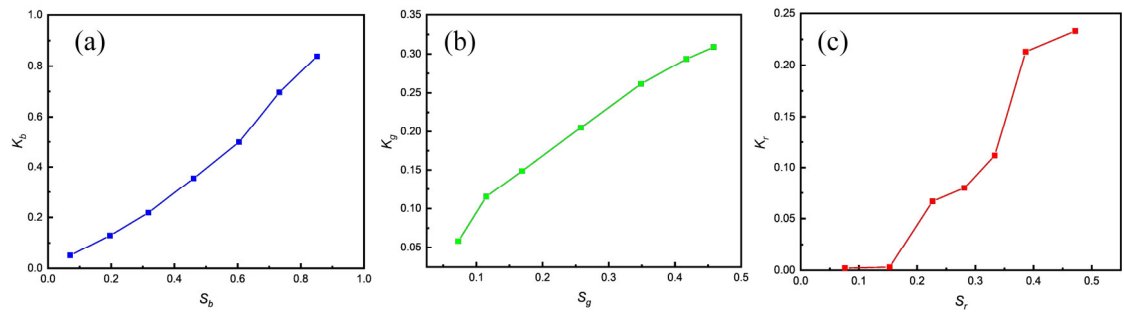

**FIG. 4.** The relative permeabilities of (a) blue fluid, (b) green fluid and (c) red fluid as a function their own saturations for the viscosity ratio of unity and the contact angles of $\theta_{bg}=45°$, $\theta_{br}=135°$ and $\theta_{gr}=135°$.

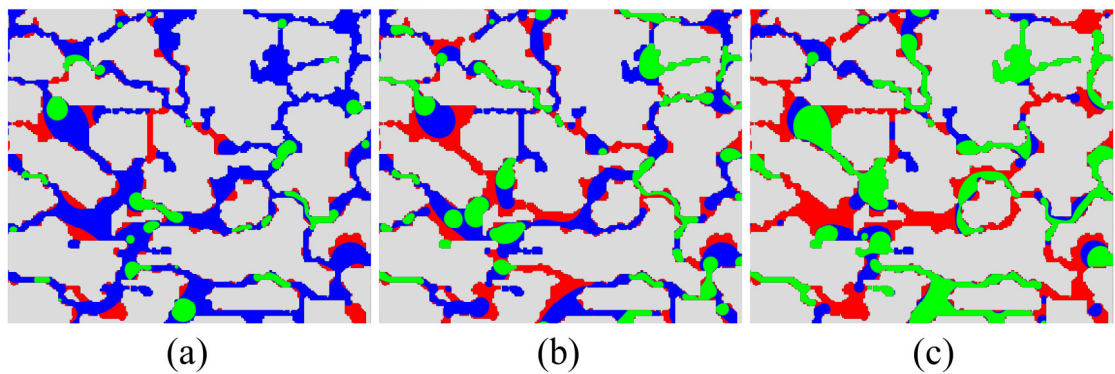

**FIG. 5.** The final fluid distributions at (a) $S_b=73.22\%$, (b) $S_b=46.09\%$ and (c)

$S_b = 19.63\%$ for the viscosity ratio of unity and the contact angles of $\theta_{bg} = 45°$, $\theta_{br} = 135°$ and $\theta_{gr} = 135°$.

Figure 6 plots three specific interfacial lengths, i.e. $L_{br}$, $L_{bg}$ and $L_{gr}$, as a function of the blue fluid saturation. Obviously, the specific interfacial length curves of $L_{br}$ and $L_{bg}$ exhibit a similar trend: As the blue fluid saturation increases, $L_{br}$ and $L_{bg}$ first increase and then decrease, with their maxima both located at $S_b = 46.09\%$. This can be explained as follows. When one fluid, e.g. the blue fluid, has a low saturation, it has few chance to contact with the other two fluids and form the interfaces. In addition, the saturations of green and red fluids are kept roughly the same, which means that $S_g$ and $S_r$ both approach 0 as $S_b$ becomes close to 1. Therefore, the specific interfacial lengths $L_{br}$ and $L_{bg}$ show a non-monotonous variation with $S_b$ and their maxima occur at moderate values of $S_b$. As $S_b$ increases, $S_g$ and $S_r$ both decrease, so the specific interfacial length between red and green fluids, i.e. $L_{gr}$, decreases monotonically, as observed in Figure 6.

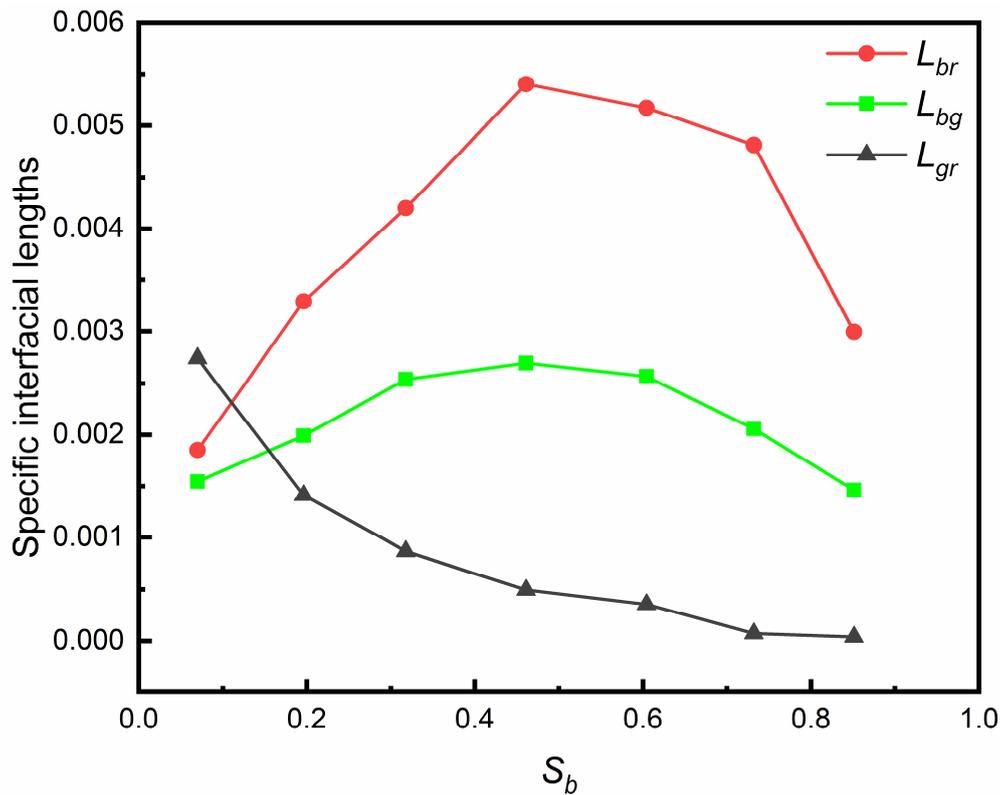

**FIG. 6.** The specific interfacial lengths $L_{br}$, $L_{bg}$ and $L_{gr}$ as a function of the blue fluid saturation for the viscosity ratio of unity and the contact angles of $\theta_{bg} = 45°$, $\theta_{br} = 135°$ and $\theta_{gr} = 135°$.

Considering that the relative permeability is not only a function of its own saturation but also dependent on the saturations of the other two fluids (Baker, 1988; Ranaee et al., 2015; Stone, 1970, 1973), we carry out simulations to investigate how the saturations of the wetting (red) and non-wetting (green) fluids influence the relative permeability of the intermediate-wetting (blue) fluid. The blue fluid saturation is fixed at $S_b = 31.77\%$, and six cases with different saturations of green and red fluids are simulated, which are achieved by adjusting the parameter $a$ whilst keeping the parameter $b$ as a constant in the initial state. All the other parameters are kept the same as those in the previous cases. Figure 7(a) shows the variation of the relative permeability of blue fluid, $K_b$, with the green fluid saturation $S_g$. It can be seen that relative permeability of blue fluid decreases with increasing $S_g$, which is attributed to the role transition of the intermediate-wetting fluid when $S_g$ increases. As shown in Figure 8, when $S_g$ is low, the pores are mainly occupied by the red and blue fluids, and the red fluid as a wetting phase prefers to occupy small pores or adhere to the rock surface forming lubricant films. As a result, the blue fluid distributed in large pores or active flow paths would experience a smaller flow resistance, leading to a larger relative permeability. However, as the $S_g$ increases, the blue fluid gradually turns its role to the wetting phase when interacting with the green fluid, so its relative permeability is lower for a higher $S_g$. On the other hand, we notice that when $S_g$ is low, the blue fluid mainly reveals clustered and continuous distribution in the void space, and thus it seldom has the chance to contact with other fluids (see Figure 8a). With increasing $S_g$, the blue fluid becomes increasingly wetting, which increases the possibility to form interfaces with the green fluid. This explains why the sum of $L_{br}$ and $L_{bg}$ shows an overall increasing trend with the increase of $S_g$, as plotted in Figure 7(b).

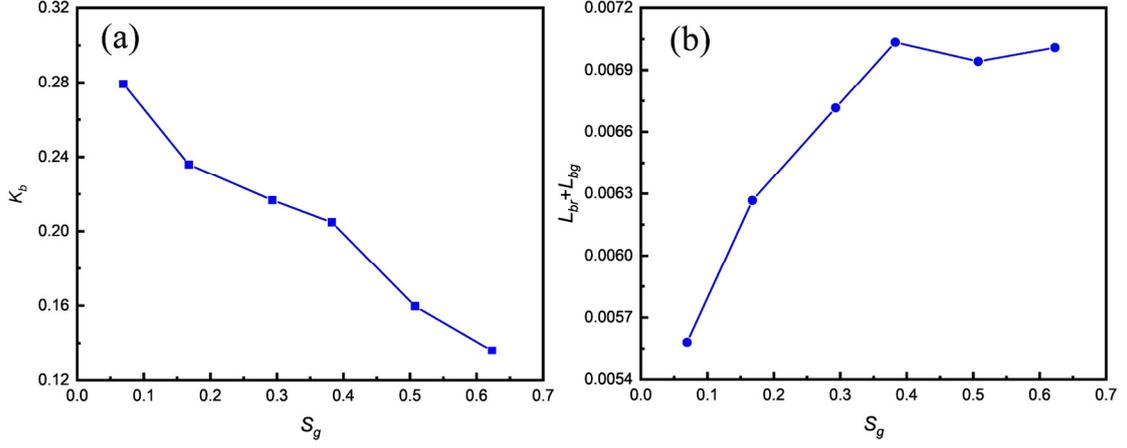

**FIG. 7.** (a) The relative permeability of blue fluid and (b) the sum of $L_{br}$ and $L_{bg}$ as a function of the green fluid saturation for $S_b = 31.77\%$, the viscosity ratio of unity and the contact angles of $\theta_{bg} = 45°$, $\theta_{br} = 135°$ and $\theta_{gr} = 135°$.

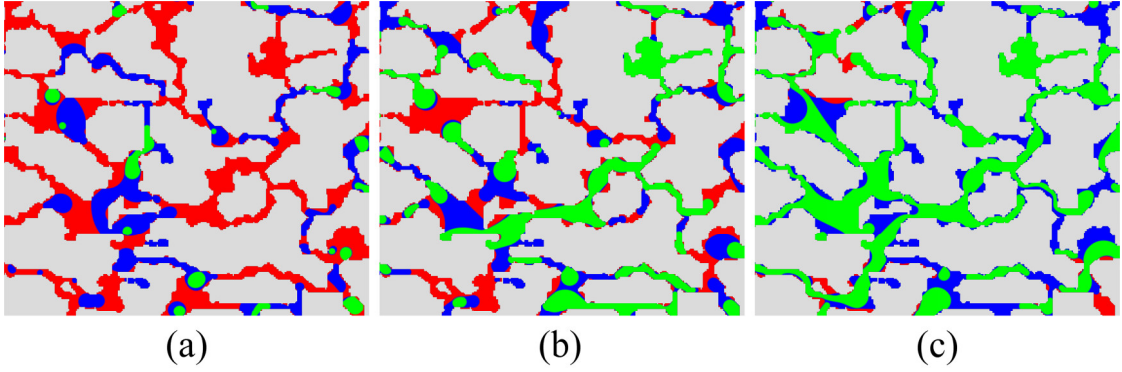

**FIG. 8.** The final fluid distributions at the green fluid saturations of (a) 6.95%, (b) 38.27% and (c) 62.30% for $S_b = 31.77\%$, the viscosity ratio of unity and the contact angles of $\theta_{bg} = 45°$, $\theta_{br} = 135°$ and $\theta_{gr} = 135°$.

*3.3. Effect of surface wettability*

Next, we investigate the effect of surface wettability on the relative permeabilities and specific interfacial lengths for the fluid saturations of $S_b = 31.77\%$, $S_g = 34.89\%$ and $S_r = 33.34\%$ and the viscosity ratio of unity. Six groups of contact angles that satisfy Eq. (15) are simulated, i.e. (a) $\theta_{bg} = 10°$, $\theta_{br} = 170°$, $\theta_{gr} = 170°$, (b) $\theta_{bg} = 30°$, $\theta_{br} = 150°$, $\theta_{gr} = 150°$, (c) $\theta_{bg} = 45°$, $\theta_{br} = 135°$, $\theta_{gr} = 135°$, (d) $\theta_{bg} = 60°$, $\theta_{br} = 120°$, $\theta_{gr} = 120°$, (e) $\theta_{bg} = 75°$, $\theta_{br} = 105°$, $\theta_{gr} = 105°$, and (f) $\theta_{bg} = 90°$, $\theta_{br} = 90°$, $\theta_{gr} = 90°$. As the contact angles vary from the group (a) to the group (f), the wetting (non-wetting) properties of the wetting (non-wetting) fluid gradually weakens, and finally three fluids become equal affinities to the rock surface. Figure 9(a) shows the relative permeabilities of three fluids as a function of the contact angle $\theta_{bg}$. Obviously, with the increase of $\theta_{bg}$, the relative permeability of the red fluid, $K_r$, monotonically increases. This is because as the red fluid becomes less wetting, the adhesive force between the red fluid and the rock surface weakens, resulting in a lower flow resistance for the red fluid. On the other hand, the relative permeabilities of the green and blue fluids, $K_b$ and $K_g$, exhibit non-monotonic variations with $\theta_{bg}$. Specifically, they first increase, then decrease to the same values when $\theta_{bg}$ increases (Figure 9a). This can be explained as follows. When $\theta_{bg}$ is low, the green fluid as a strong non-wetting phase would experience a small adhesive force from the rock surface, but the red fluid is affiliative to the rock surface so that it permeates into small pores and blocks them, forming many inactive regions as circled by black rectangles in Figure 10(a). This would reduce the mobility of both green and blue fluids, leading to low permeabilities for these two fluids. As $\theta_{bg}$ increases, the mobility of red fluid strengthens, and the blockage of narrow pore spaces is mitigated. Thus, the green and blue fluids reveal higher relative permeabilities. Upon further increasing the value of $\theta_{bg}$, both green and blue fluids get less non-wetting and this effect becomes a dominant factor. As a result, $K_b$ and $K_g$ decrease, and finally they reach a similar value to $K_r$ at $\theta_{bg} = 90°$, which is reasonable because all three fluids are of equal affinity to the solid surface.

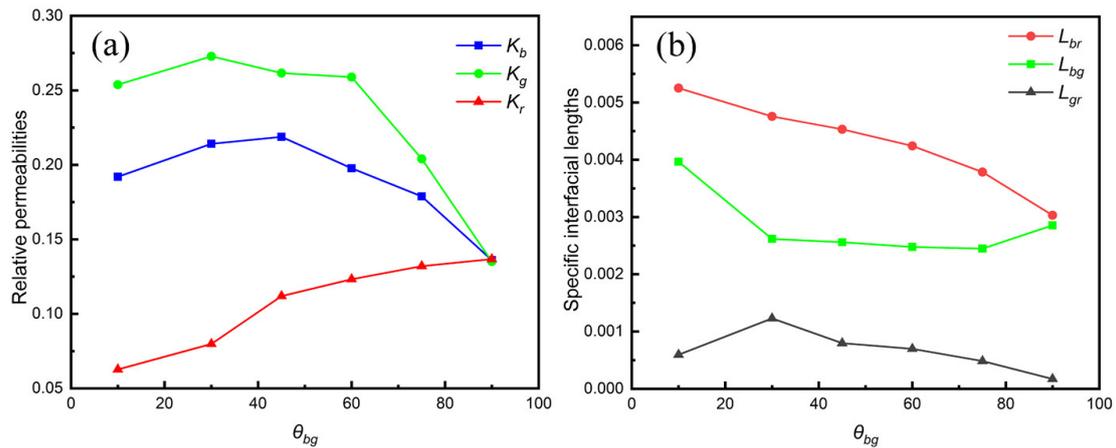

**FIG. 9.** (a) The relative permeabilities of three fluids and (b) the specific interfacial lengths as a function of the contact angle $\theta_{bg}$ for the viscosity ratio of unity and the saturations of $S_b = 31.77\%$, $S_g = 34.89\%$ and $S_r = 33.34\%$.

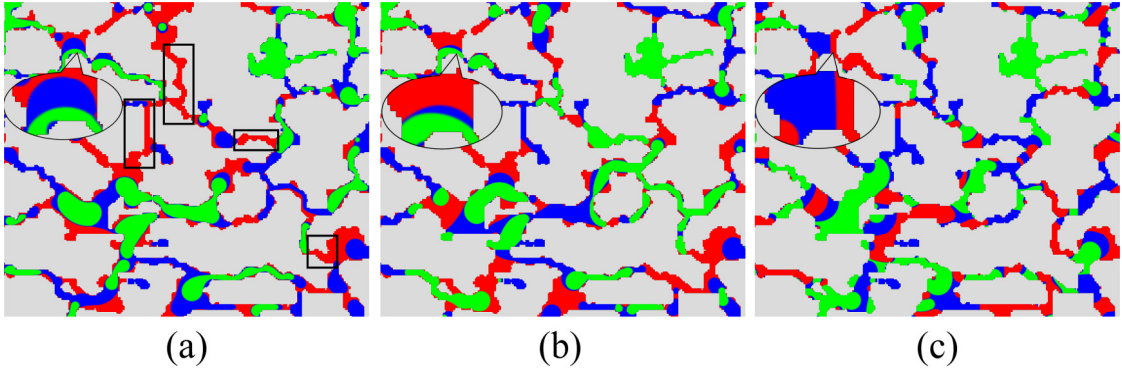

(a) (b) (c)

**FIG. 10.** The final fluid distributions at the contact angles of (a) $\theta_{bg} = 10°$, $\theta_{br} = 170°$, $\theta_{gr} = 170°$, (b) $\theta_{bg} = 45°$, $\theta_{br} = 135°$, $\theta_{gr} = 135°$ and (c) $\theta_{bg} = 90°$, $\theta_{br} = 90°$, $\theta_{gr} = 90°$ for the viscosity ratio of unity and the saturations of $S_b = 31.77\%$, $S_g = 34.89\%$ and $S_r = 33.34\%$. In (a), the black rectangles are to highlight the inactive regions blocked by the strong wetting (red) fluid. For all the subfigures, the same regions are zoomed in (ellipse border lines) to clearly show the difference of the phase interfaces.

Figure 9(b) shows the interfacial lengths versus the contact angle $\theta_{bg}$. It can be seen that with increasing $\theta_{bg}$, $L_{br}$ decreases monotonously and overall exhibits a tendency to converge toward $L_{bg}$. Compared with $L_{br}$ and $L_{bg}$, $L_{gr}$ does not change much with $\theta_{bg}$ and remains at relatively low values. These behaviors could be explained as follows. When $\theta_{bg}$ is low, the red fluid as a strong wetting fluid would adhere to the rock surface forming a long and narrow lubricant film, as shown in Figure 10(a). Thus, there is a long phase interface between red fluid and other fluids. However, considering the values of three interfacial tensions, i.e. $\sigma_{gr}:\sigma_{bg}:\sigma_{br}=2:1:1$, to minimize the surface energy, the red fluid prefers not to contact with the green fluid, and the blue fluid would form a layer between the green and red fluids, which is known as spreading layer (Alizadeh and Piri, 2015; Jiang and Tsuji, 2016). Therefore, $L_{gr}$ does not vary too much but stabilizes at a lower value compared with $L_{br}$ and $L_{bg}$ for all the wetting conditions. With the increase of $\theta_{bg}$, the red fluid gradually becomes less wetting, and its influence is: (1) The lubricant film adhering to the rock surface is reduced, and thus $L_{br}$ decreases, especially along the flow paths; (2) The curvature of the phase interface between the blue and red fluids decreases, leading to a reduced interfacial length in pore bodies and throats. The latter can be clearly seen from the local enlarged view in Figure 10, where the interface curvature between the blue and red fluids is greater when $\theta_{bg}$ is lower. Thus, it is expected that $L_{br}$ decreases with $\theta_{bg}$. When $\theta_{bg}$ increases to $90°$, the green and blue fluids are of the same wettability, so the $L_{br}$ and $L_{bg}$ curves intersect at this point.

*3.4. Effect of viscosity ratios*

Finally, the effect of the viscosity ratios is investigated for the contact angles of $\theta_{bg}=45°$, $\theta_{br}=135°$, $\theta_{gr}=135°$ and the fluid saturations of $S_b=31.77\%$, $S_g=34.89\%$, $S_r=33.34\%$. Two viscosity ratios of $\mu_g/\mu_b$ and $\mu_r/\mu_b$ are considered here, and when the effect of $\mu_g/\mu_b$ ($\mu_r/\mu_b$) is investigated, the other viscosity ratio $\mu_r/\mu_b$ ($\mu_g/\mu_b$) is fixed at unity. The simulated viscosity ratios range from 0.025 to 40, far beyond those achieved by the existing LBM studies in which the viscosity ratio was limited to at most 5 (Jiang and Tsuji, 2016; Zhang et al., 2019).

Figure 11(a) plots the relative permeabilities of three fluids as a function of the viscosity ratio. It can be seen that $K_g$ ($K_r$) increases but $K_r$ and $K_b$ ($K_g$ and $K_b$) decrease with the viscosity ratio of $\mu_g/\mu_b$ ($\mu_r/\mu_b$), which can be explained as follows. Without loss of generality, we only analyze the case in which $\mu_g/\mu_b$ is varied. With the increase of $\mu_g/\mu_b$, the flow resistance exerted on the green fluid by the other two fluids decreases due to the viscous coupling effect (Jiang et al., 2020), leading to an enhanced capability to flow through the porous media for the green fluid. On the contrary, the flow resistance exerted on the red or blue fluid increases as $\mu_g/\mu_b$ increases. Therefore, the values of $K_r$ and $K_b$ both decrease with increasing $\mu_g/\mu_b$. We also notice that three relative permeabilities are not equal when three fluids have the same viscosities. This is attributed to the effect of the surface wettability: the fluid affiliative to the rock surface would undergo a larger adhesive force from the rock surface and thus shows a lower permeability, as discussed in the previous sections.

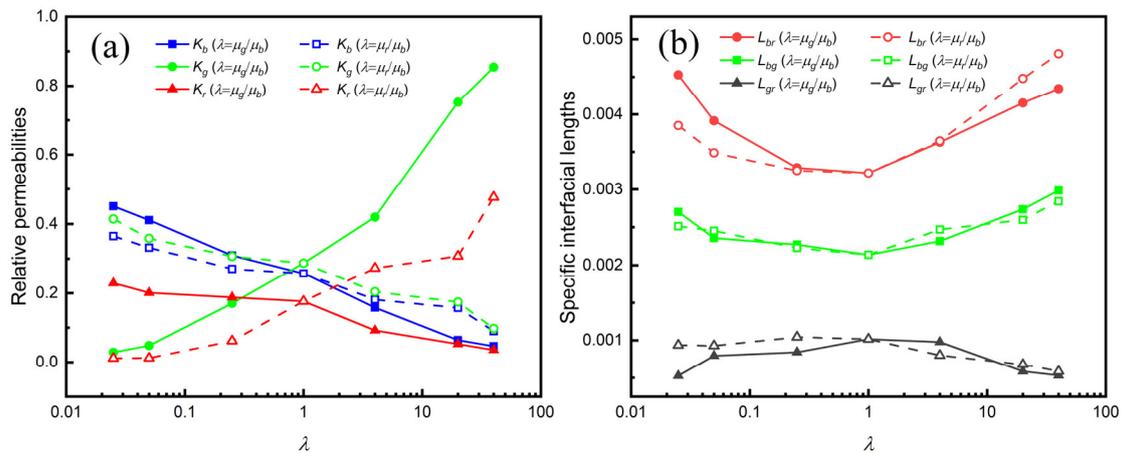

**FIG. 11.** (a) The relative permeabilities of three fluids and (b) the specific interfacial lengths as a function of the viscosity ratio for the contact angles of $\theta_{bg}=45°$,

$\theta_{br} = 135°$, $\theta_{gr} = 135°$ and the fluid saturations of $S_b = 31.77\%$, $S_g = 34.89\%$, $S_r = 33.34\%$.

Figure 11(b) shows the effect of the viscosity ratios on the specific interfacial lengths. Obviously, as the viscosity ratio $\lambda$ increases, the specific interfacial lengths $L_{br}$ and $L_{bg}$ first decrease and then increase, with their minima occurring at $\lambda = 1$. This trend can be explained as a result of unstable fingering. The greater the viscosity ratio deviates from unity, the more unstable the flows become, which would lead to an increase in the specific interfacial length. The above explanation can be justified by the velocity distributions shown in Figures 12 and 13. It is clearly seen that the velocity field is more non-uniform when the viscosity ratio deviates from 1, which indicates that the flows are more unstable under this condition. Such instability was also observed in the experiment (Bischofberger et al., 2014), which is related to the generation of viscous fingering patterns when the viscosity ratio is high. Note that a similar phenomenon was also reported in two-phase seepage problem by Xu et al. (Xu and Liu, 2018). Compared with the specific interfacial lengths $L_{br}$ and $L_{bg}$, the specific interfacial length $L_{gr}$ presents much lower values, which is due to the presence of spreading layer between the green and red fluids (Alizadeh and Piri, 2015; Jiang and Tsuji, 2016) for the used interfacial tensions ($\sigma_{gr}:\sigma_{bg}:\sigma_{br} = 2:1:1$).

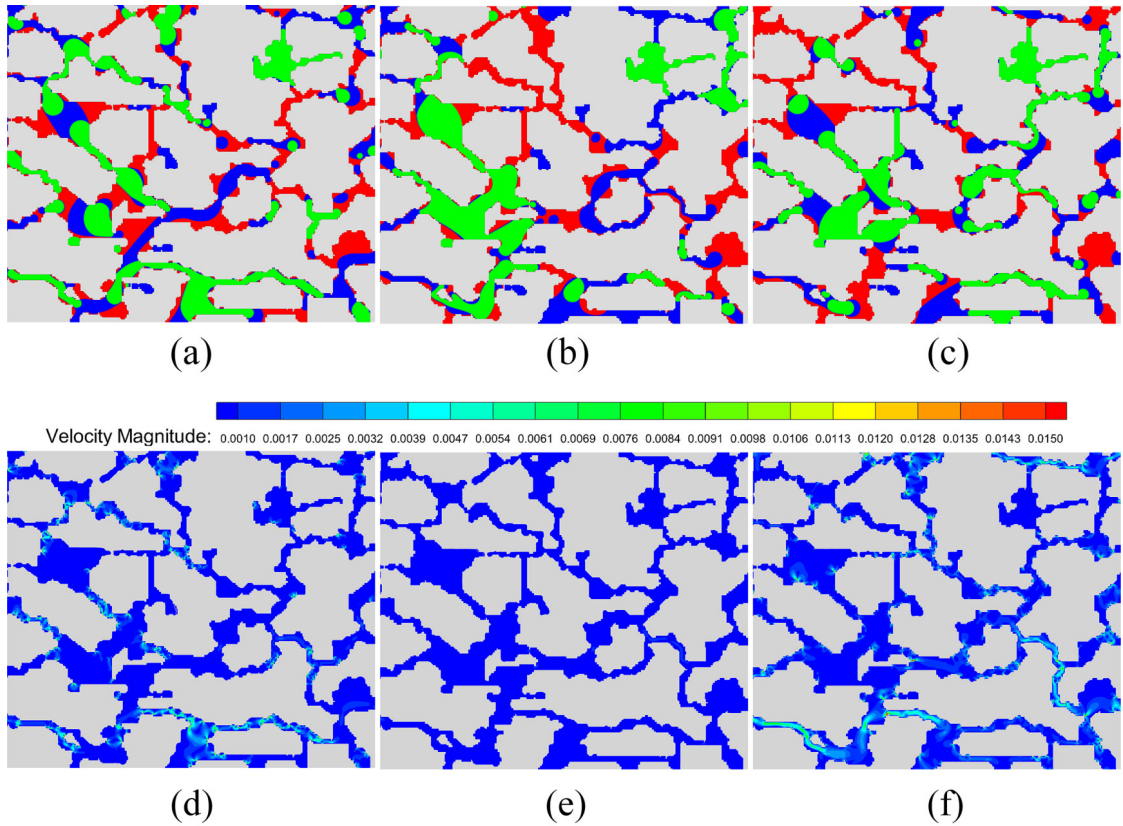

**FIG. 12.** The final fluid distributions at the viscosity ratios of (a) $\mu_g / \mu_b = 0.025$, (b) $\mu_g / \mu_b = 1$ and (c) $\mu_g / \mu_b = 40$ for the contact angles of $\theta_{bg} = 45°$, $\theta_{br} = 135°$, $\theta_{gr} = 135°$ and the fluid saturations of $S_b = 31.77\%$, $S_g = 34.89\%$, $S_r = 33.34\%$. Note that the velocity distributions corresponding to (a)-(c) are shown in (d)-(f).

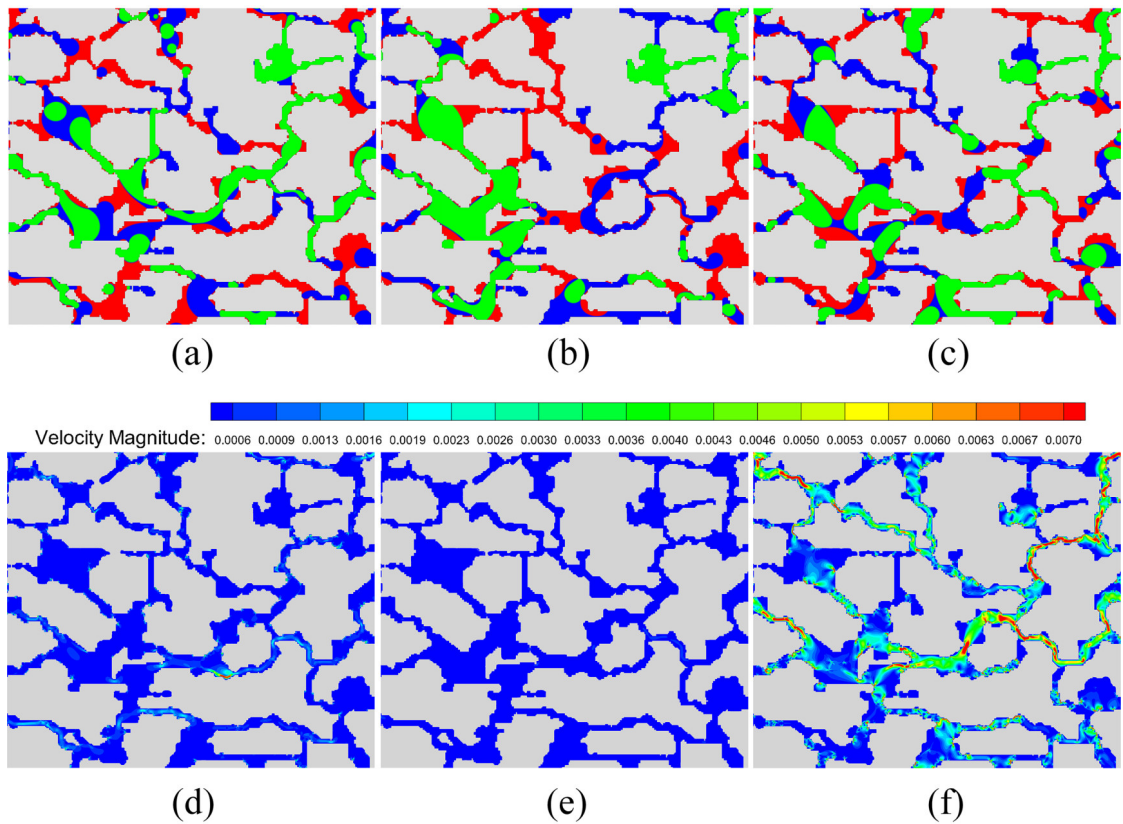

**FIG. 13.** The final fluid distributions at the viscosity ratios of (a) $\mu_r/\mu_b = 0.025$, (b) $\mu_r/\mu_b = 1$ and (c) $\mu_r/\mu_b = 40$ for the contact angles of $\theta_{bg} = 45°$, $\theta_{br} = 135°$, $\theta_{gr} = 135°$ and the fluid saturations of $S_b = 31.77\%$, $S_g = 34.89\%$, $S_r = 33.34\%$. Note that the velocity distributions corresponding to (a)-(c) are shown in (d)-(f).

## 4. Conclusions

In this work, a lattice Boltzmann color-gradient model recently improved by our group (Li et al., 2021; Yu et al., 2019b) is used to simulate immiscible three-phase flows in a 2D pore network extracted from a slice of realistic Berea sandstone. Different from the previous studies, the present work focuses on a ternary fluid system where three interfacial tensions do not satisfy the Neumann's triangle. A constant body force is applied to drive the fluid movement, and the effect of fluid saturations, surface wettability and viscosity ratios on the relative permeabilities and specific interfacial lengths is systematically investigated. The major conclusions are summarized as follows.

The relative permeability of each fluid increases as its own saturation increases. Two specific interfacial lengths, namely the one between intermediate wetting and non-wetting fluids and the one between intermediate wetting and wetting fluids, first increase and then decrease with the saturation of intermediate wetting fluid, while the specific interfacial length between wetting and non-wetting fluids decreases monotonously. For constant saturations of three fluids, the variation of the surface wettability would change the adhesive force between fluids and rock surface, thus influencing the relative permeabilities. The relative permeability of wetting fluid increases as the wetting (non-wetting) fluid becomes less wetting (non-wetting), while the relative permeabilities of the other two fluids show a non-monotonous variation. Due to the presence of spreading layer (consisting of intermediate wetting fluid), the specific interfacial length between wetting and non-wetting fluids stabilizes at a lower level at all times, compared with the other two specific interfacial lengths. In addition, increasing the viscosity ratio of wetting (non-wetting) to intermediate wetting fluids increases the relative permeability of wetting (non-wetting) fluid but decreases the other two relative permeabilities. As the viscosity ratio deviates more from 1, the non-uniformity of the velocity distribution increases and the phase interfaces become more unstable, thus increasing the specific interfacial length.

## Acknowledgments

The authors acknowledge the financial support of the National Key Research and Development Program of China (No. 2018YFA0702400), the National Natural Science Foundation of China (Grant Nos. 12072257, 51876170), the National Key Project (Grant No. GJXM92579), and JSPS through a Grant-in-Aid for Young Scientists (Grant No. 19K15100).

immiscible fluids in random porous media determined by the lattice Boltzmann method. International Journal of Heat and Mass Transfer 134, 311-320.